\documentclass[journal]{IEEEtran}
\ifCLASSINFOpdf
   \usepackage[pdftex]{graphicx}
\else
  \usepackage[dvips]{graphicx}
\fi

\usepackage{multirow}
\usepackage[cmex10]{amsmath}
\usepackage{amssymb}
\usepackage{color}
\usepackage{mathtools}
\usepackage{algorithmic}

\usepackage{array}
\usepackage{xparse}

\usepackage{mdwmath}
\usepackage{mdwtab}

\usepackage{listings}
\usepackage{color} 
\definecolor{mygreen}{RGB}{28,172,0} 
\definecolor{mylilas}{RGB}{170,55,241}

\begin{document}

\title{An Efficient Analytical Evaluation of the Electromagnetic Cross-Correlation Green's Function in MIMO Systems}

\author{Debdeep Sarkar, \textit{Member, IEEE}, Said Mikki, \textit{Senior Member, IEEE}, Yahia Antar, \textit{Life Fellow, IEEE}}

\maketitle

\footnotetext[1]{Debdeep Sarkar \textit{(Corresponding Author)} and Yahia Antar are with the Royal Military College, PO Box 17000, Station Forces Kingston, Ontario, K7K 7B4, Canada (Emails: \textit{debdeep1989@gmail.com}; antar-y@rmc.ca).}
\footnotetext[2]{Said Mikki is with University of New Haven, West Haven, Connecticut, 300 Boston Post Rd, 06516, USA (Email: said.m.mikki@gmail.com).}
\footnotetext[3]{$\copyright$20XX IEEE.  Personal use of this material is permitted.  Permission from IEEE must be obtained for all other uses, in any current or future media, including reprinting/republishing this material for advertising or promotional purposes, creating new collective works, for resale or redistribution to servers or lists, or reuse of any copyrighted component of this work in other works.”}





\begin{abstract}
\boldmath
In this paper, we completely eliminate all numerical integrations needed to compute the far-field envelope cross-correlation (ECC) in multiple-input-multiple-output (MIMO) systems by deriving accurate and efficient analytical expressions for the frequency-domain cross-correlation Green's functions (CGF), the most fundamental electromagnetic kernel needed for understanding and estimating spatial correlation metrics in multiple-antenna configurations. The analytical CGF is derived for the most general three-dimensional case, which can be used for fast CGF-based correlation matrix calculations in MIMO systems valid for arbitrary locations and relative polarizations of the constituent elements.  
\end{abstract}

\noindent




%
\IEEEpeerreviewmaketitle

\section{Introduction}
Fifth generation (5G) wireless networks aiming at high data-rate ($>10$ Gbps) and efficient interference suppression between multiple users in ultra-dense networks (UDNs), deploy large antennas arrays/massive multiple-input-multiple-output (MIMO) systems as key enabling technology \cite{mimo_book1}-\cite{mimo_book4}. One crucial aspect of such multiple-antenna systems in MIMO transceivers is the latter's ``spatial correlation matrix'', which accounts for mutual interaction (conventionally, only far-field is taken into account) between the constituent antenna element-pairs \cite{spacetime}-\cite{ampoma_1}. Despite the pivotal role of this spatial correlation matrix in shaping the overall channel matrix and consequent capacity/interference-suppression issues (see \cite{jensen_wallace}-\cite{janaswamy} for details), the impact of pure antenna effects and various core electromagnetic aspects in MIMO channel modelling are often not emphasized adequately. 

Traditionally, this antenna spatial correlation performance is determined from a formula involving the radiation patterns of individual elements \cite{karaboikis}, which makes it very cumbersome to perform antenna current level optimization aiming at a desired diversity performance. To relate the antenna correlation directly to the radiating antenna \textit{current} distribution (i.e., essentially bypassing altogether the original far-field pattern route), the concept of cross-correlation Green's functions (CGFs) was first introduced in \cite{mikki_cgf_2015},\cite{mikki_book} and later elaborated for antenna design applications \cite{clauzier2}, \cite{mikki_access}. A preliminary step of this  CGF-based correlation calculation is to construct a suitable infinitesimal dipole model (IDM) for the radiating MIMO antenna current distribution (see \cite{mikki_idm1}-\cite{yang_idm_2} for detailed theory and applications of IDM). The next step is to employ suitable CGFs in order to compute individual ID-pair interactions systematically and then combine them togethoer to construct the global correlation matrices of the system \cite{mikki_cgf_2015}. Application of CGFs to realize high diversity gain MIMO antenna arrays as well as dual-polarized massive MIMO systems has been reported in recent past \cite{clauzier1}-\cite{idm_cgf_mmimo_2019}. By efficient integration of the CGFs with finite-difference-time-domain (FDTD) computational paradigm, one can also perform wide-band time-domain correlation analysis for arbitrary antennas \cite{ds_tap_1}-\cite{ds_tap_2}. The CGFs are also extended to deal with radiators involving both electric/magnetic current sources \cite{ds_aps_2018}. Possible application of the CGF methodology for near-field stochastic systems \cite{mikki_stochastic} and antenna directivity analysis \cite{directivity_cgf} are also being actively explored presently. 

However, calculation of the CGF tensor components requires numerical integration involving elevation ($\theta$) and azimuth ($\phi$) angle dependent terms in the argument of complex exponential functions \cite{mikki_cgf_2015}, \cite{ds_tap_1}. Therefore, it becomes extremely difficult to efficiently embed these CGFs in fast optimization routines aiming at finding optimum current distributions for desired diversity performance. This necessitates a robust analytical evaluation scheme for the determination of the CGF tensor components and the possibility of achieving this was in fact already suggested in \cite{mikki_cgf_2015}. Although some specialized approximation formulas of the time-domain CGFs were attempted in \cite{ds_eucap1}, IDM-synthesis of MIMO antennas strictly requires frequency-domain CGFs, and the analytical evaluation of the latter at a very general level has been so far an open problem. Mitigating this shortcoming in the literature will be the main contribution of the present work.  

In this paper, we first employ a series-expansion approach to approximate the complex exponential functions in CGF tensors (some preliminary ideas were briefly suggested \cite{ds_aps2019_1}). Next, by deploying carefully selected mathematical properties enjoyed by the Beta and Gamma functions, we analytically evaluate the full angular space integration involving oscillatory terms. In this way, analytical expressions of CGF tensors are presented here for the most general three-dimensional case. 



\section{Analytical CGF Determination in One/Two/Three Dimensional Arrays}

\subsection{Review of the CGF Tensor}

In standard MIMO literature, the complex correlation coefficient $\rho$ between the far-field patterns $\mathbf{E}_{1}=\mathbf{E}_{1}\left(\theta,\phi\right)$ and $\mathbf{E}_{2}=\mathbf{E}_{2}\left(\theta,\phi\right)$, respectively generated by complex current distributions $\mathbf{J}_{1}=\mathbf{J}_{1}\left(\mathbf{r'}\right)$ and $\mathbf{J}_{2}\left(\mathbf{r''}\right)$, is traditionally calculated by \cite{karaboikis} 
\begin{equation} \label{rho_gen}
    \rho=\frac{\int_{4\pi} \left[\mathbf{E}_{1}\cdot \mathbf{E}_{2}^{*}\right] d\Omega}{\sqrt{\int_{4\pi}  \left[\mathbf{E}_{1}\cdot \mathbf{E}_{1}^{*}\right] d\Omega \int_{4\pi} \left[\mathbf{E}_{1}\cdot \mathbf{E}_{1}^{*}\right] d\Omega}},
\end{equation}
where $d\Omega$ is the solid angle element given by $d\Omega=\sin\theta d\theta d\phi$. That is, only the radiation fields appear in the original definition. This makes the process of evaluating $\rho$ and desinging optimum antennas for spatial diversity applications very challenging since the 3D computation of the far-field pattern is demanding. Moreover, the geometrical details of the radiator, e.g., shape, orientations, excitations, do not directly manifest themselves in the far field. For those reasons, in \cite{mikki_cgf_2015} $\rho$ is expressed directly in terms of $\mathbf{J}_{1}$ and $\mathbf{J}_{2}$ as:
\begin{equation} \label{rho_cgf}
    \rho=\frac{\int d^{3}\mathbf{r'} \int d^{3}\mathbf{r''}   \mathbf{J}_{1}\cdot \mathbf{\Bar{C}} \cdot \mathbf{J}^{*}_{2}}{\sqrt{\left[ \int d^{3}\mathbf{r'} \int d^{3}\mathbf{r''}  \mathbf{J}_{1}\cdot \mathbf{\Bar{C}} \cdot \mathbf{J}^{*}_{1} \right]\left[ \int d^{3}\mathbf{r'} \int d^{3}\mathbf{r''}  \mathbf{J}_{2}\cdot \mathbf{\Bar{C}} \cdot \mathbf{J}^{*}_{2} \right]}},
\end{equation}
where all integrals are performed over the entire antenna radiating surface (the support of the current distribution functions $\textbf{J}_1(\textbf{r})$ and $\textbf{J}_2(\textbf{r})$.
Here, $\mathbf{\Bar{C}}=\mathbf{\Bar{C}}\left(\mathbf{r'},\mathbf{r''} \right)$ stands for the CGF tensor in uniform propagation environment given by \cite{mikki_cgf_2015}:
\begin{equation} \label{cgf_mod}
    \mathbf{\Bar{C}}\left(\mathbf{r'},\mathbf{r''} \right)= \int_{0}^{\pi} \int_{0}^{2\pi} \left[\mathbf{\Bar{I}}-\hat{r}\hat{r} \right] e^{j\mathbf{k}\cdot\left(\mathbf{r'}-\mathbf{r''}\right)} \sin\theta d\theta d\phi. 
\end{equation}
The quantity $\mathbf{\bar I}$ is the unit dyad, while $\mathbf{r'}$ and $\mathbf{r''}$ denote the spatial dependencies of $\mathbf{J}_{1}$ and $\mathbf{J}_{2}$, respectively. Moreover, $\mathbf{k}=k\hat{r}$, with $k=2\pi/\lambda$ ($\lambda=$ operating wavelength) and $\hat{r}$ being the radial unit-vector in spherical coordinate system given by
\begin{equation}\label{radial position}
   \hat r\left(\theta ,\varphi\right)=\hat r\left(\Omega\right)  := \hat x\cos \varphi \sin \theta  + \hat y\sin \varphi \sin \theta  + \hat z\cos \theta.  
\end{equation}
The nine components $C_{pq}$ (where $p=x,y,z$ and $q=x,y,z$) of the CGF tensor $\mathbf{\Bar{C}}\left(\mathbf{r'},\mathbf{r''} \right)$ in \eqref{cgf_mod} can be derived after using \eqref{radial position} and elementary dyadic arithmetic rule. The results are expressed as:
\begin{equation} \label{cpq_exact}
    C_{pq}=\int_{0}^{\pi} \int_{0}^{2\pi} f_{pq} \exp \left[jkr_{d} \right] d\theta d\phi,    
\end{equation}
where $r_{d}=\hat{r}\cdot\left(\mathbf{r'}-\mathbf{r''}\right)=x_{d}\sin\theta \cos\phi + y_{d}\sin\theta\sin\phi + z_{d}\cos\theta$, with $x_{d}=x'-x''$, $y_{d}=y'-y''$ and $z_{d}=z'-z''$ and values of $f_{pq}=f_{pq}\left(\theta,\phi\right)$ are defined as \cite{mikki_cgf_2015}, \cite{ds_tap_1}:
\begin{eqnarray} \label{fpq}
f_{xx} &=& \sin\theta \left(1-\sin^{2}\theta\cos^{2}\phi \right) \nonumber, \\
f_{yy} &=& \sin\theta \left(1-\sin^{2}\theta\sin^{2}\phi \right) \nonumber, \\
f_{zz} &=& \sin^{3}\theta, \, \\
f_{xy} = f_{yx} &=& -\sin^{3}\theta, \cos\phi \sin\phi \nonumber, \\
f_{yz} = f_{zy} &=& -\sin^{2}\theta \cos\theta \sin\phi \nonumber, \\
f_{zx} = f_{xz} &=& -\sin^{2}\theta \cos\theta \cos\phi. 
\end{eqnarray}
Consequently, the CGF-based technique completely eliminates the requirement of going through the conventional pattern-based route of \eqref{rho_gen} by focusing instead on the total radiating antenna \textit{current} distribution, i.e., only current points reflecting the radiator excitation and geometry are needed, and these are considerably smaller in number than the far-field points \cite{mikki_cgf_2015}, \cite{ds_tap_1}, \cite{ds_tap_2}.
Clearly, still all components of the CGF involve two-dimensional angular integration operations over the entire sphere. In general, for every position pair $\textbf{r}',\textbf{r}''$, these integrals must be computed again. Therefore, for large antenna arrays the net number of numerical computations becomes large. 

\subsection{General Idea of CGF Tensor Approximation}

While evaluation of $C_{pq}$ via \eqref{cpq_exact} requires computing an angular-space integration of $f_{pq}\exp{\left(jkr_{d}\right)}$, it was pointed out in \cite{mikki_cgf_2015} that CGF might be computed or at least well-approximated analytically. Various approaches may be pursued here. For example, it is possible to expand the integrand of every integral into orthogonal functions then evaluate angular integrations using exact orthogonality relations. However, the orthogonal expansion itself often requires numerical integrations to obtain the needed Fourier coefficients (the weight of every orthogonal function) and hence it may not lead to efficient algorithm. Moreover, because each of the nine integrals entering into the determination of the full $3 \times 3$ dyad $\mathbf{\Bar{C}}$ may involve a distinct angular function in the integrand, performing an orthogonal function expansion here becomes cumbersome and very tedious.

Another idea is to simplify \eqref{cpq_exact} by making use of a Bessel's function utilizing the Jacobi-Anger expansion \cite{beta_gamma}. However, an alternative approach is proposed in this paper, where we simply utilize the familiar Taylor series expansion of the exponential functions 
\begin{equation}
    \exp \left[jkr_{d} \right] = \sum_{n=0}^{\infty} \frac{\left(jkr_{d}\right)^{n}}{n!},
\end{equation}
after which $C_{pq}$ in \eqref{cpq_exact} can be approximated by truncating into finite number of $N$ terms giving 
\begin{equation} \label{cpq_approx}
    C_{pq} \approx \sum_{n=0}^{N}  \left( jk \right)^{n} I_{pq,n},   
\end{equation}
where,
\begin{equation} \label{ipq_0}
    I_{pq,n}= \frac{1}{n!} \int_{0}^{\pi} \int_{0}^{2\pi} f_{pq} \left( r_{d}. \right)^{n} d\theta d\phi.  
\end{equation}
The subsequent sections will build fast and robust algorithm for the evaluation of the quantity $I_{pq,n}$ for various infinitesimal dipole (ID) array configurations (one/two/three dimensional topologies will be considered). 

We also note at this juncture that the formal limit $N \rightarrow \infty$ in \eqref{cpq_approx} will give the \textit{exact} $C_{pq}$ value as obtained from \eqref{cpq_exact}. However, soon we will discover that opting for $N \rightarrow \infty$ is not required in our correlation matrix computation algorithm. The proposed method then provides a tradeoff between exactness and computational efficiency. While the use of $N$ makes our evaluation less exact than using complete orthogonal function series expansion, nevertheless the final algorithm turns out to be very efficient for reasonably finite values of $N$, while it avoids the mathematical complexity of the angular spherical function approach. 

\begin{figure} [htbp]
\begin{center}
\includegraphics[width=7.5cm]{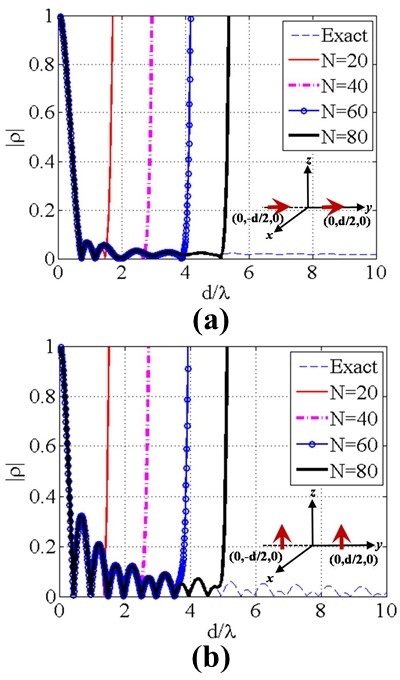}
\caption{Variation of $|\rho|$ with respect to normalized inter-element spacing $p=d/\lambda$, for: (a) two $y$-directed IDs, and (b) two $z$-directed IDs, placed along $y$-axis separated $y_{d}=d$ (schematics shown in inset). We perform numerical calculation using the exact formula \eqref{cpq_exact} and the series-approximation formula \eqref{cpq_approx} for $N=20,40,60,80$.} \label{fig1}
\end{center}
\vspace{-10pt}
\end{figure}
 
\subsection{Analytical CGF Estimation for One-dimensional Linear Dipole Arrays}

To begin with, let us consider the linear/one-dimensional array configuration that consists of infinitesimal dipoles (IDs) strictly placed along the $y$-axis, but yet are allowed to have arbitrary polarization and inter-element spacing. Since $x_{d}=z_{d}=0$ for this case, we have $r_{d}= y_{d}\sin\theta\sin\phi$. Applying this $r_{d}$ in \eqref{ipq_1}, the reduced expression for $I_{pq,n}$ is obtained as follows:
\begin{equation} \label{ipq_1}
    I_{pq,n}= \frac{\left(y_{d} \right)^{n}}{n!} G_{pq,n},   
\end{equation}
where,
\begin{equation} \label{gpq_1}
    G_{pq,n}=\int_{0}^{\pi} \int_{0}^{2\pi} f_{pq} \left( \sin\theta \sin\phi \right)^{n} d\theta d\phi.
\end{equation}
Now the natural question that arises here is this: \textit{how can one decide the proper value $n=N$ in \eqref{cpq_approx} needed to efficiently compute the far-field correlation of given ID pair with acceptable accuracy?}

To answer this question, we probe further into $C_{pq}$, by expressing the inter-element spacing $y_{d}$ in terms of the operating wavelength $\lambda$. Writing $y_{d}=p\lambda$ and using $k=2\pi/\lambda$, one can put $C_{pq}$ via \eqref{cpq_approx}, \eqref{ipq_1} and \eqref{gpq_1} in the following form 
\begin{equation} \label{cpq_approx_1D_1}
    C_{pq} \approx \sum_{n=0}^{N}  j^{n} G_{pq,n} \left[\frac{\left( 2\pi p \right)^{n}}{n!} \right].     
\end{equation}
At this point, we use the well known Stirling's formula for $n!$, which is very accurate for large values of $n$ \cite{dutka}-\cite{robbins}:
\begin{equation} \label{stirling}
    n!=\sqrt{2\pi n} \left( \frac{n}{e} \right)^{n}.
\end{equation}
This further reduces \eqref{cpq_approx_1D_1} to:
\begin{equation} \label{cpq_approx_1D}
    C_{pq} \approx \sum_{n=0}^{N} j^{n} \left( \frac{G_{pq,n}}{\sqrt{2\pi n}} \right) \left( \frac{2\pi e p}{n} \right)^{n}.
\end{equation}
Note that, the factor $G_{pq,n}/\sqrt{2\pi n}$ in \eqref{cpq_approx_1D} decreases asymptotically with $n$, and is ignored for the time-being. By careful observation of the next term involving $n$-th power of $(2\pi e p/n)$ and using $2\pi e \approx 17.08$, we provide a \textit{rule-of-thumb} to determine the necessary $n=N$ for a given value of $p$ required to truncate the series in \eqref{cpq_approx} yet while not compromising correlation calculation accuracy:
\begin{equation} \label{ruleofthumb}
     N > 17p.
\end{equation}
For the example $p=5$, i.e. if the two IDs are placed $5\lambda$ apart, one should need approximately 85 terms to accurately determine $C_{pq}$ from \eqref{cpq_approx}. A numerical verification of this rule-of-thumb formula \eqref{ruleofthumb} can be found in Fig. \ref{fig1}(a) and Fig. \ref{fig1}(b), where variations of $|\rho|$ with respect to inter-element spacing $p=d/\lambda$ (where $y_{d}=d$) are shown respectively for a $y$-directed and $z$-directed ID pair, placed along the $y$-axis. One can observe from Fig. \ref{fig1}(a) and Fig. \ref{fig1}(b) that for $d/\lambda > N/17$, the $|\rho|$ value using \eqref{cpq_approx} quickly deviates from that determined via the exact formula \eqref{cpq_exact}. Observing Fig. \ref{fig1}(a) and Fig. \ref{fig1}(b), it can be said that for inter-element spacing $>6 \lambda$, spatial correlation magnitude $|\rho|$ is sufficiently small ($|\rho|<0.1$), and may be ignored for practical application purpose. This fact will also come in handy in formulating a general spatial correlation determination algorithm in the subsequent section.    


The next challenge is to determine $G_{pq,n}$ analytically, therefore completely eliminating the need for numerical integration routines, as emphasized before. We demonstrate the derivation for $G_{zz,n}$ to start with. Using $f_{zz}$ from \eqref{fpq} in \eqref{gpq_1}, we obtain:
\begin{multline} \label{gzz_n}
    G_{zz,n}=\int_{0}^{\pi} \int_{0}^{2\pi} f_{zz} \left( \sin\theta \sin\phi \right)^{n} d\theta d\phi \\= \left[ \int_{0}^{\pi} \sin^{n+3}\theta d\theta \right] \left[ \int_{0}^{2\pi} \sin^{n}\phi d\phi \right].
\end{multline}
Note that, $n$ can be either odd or even. For \textit{odd} values of $n$, the integral with $\phi$-dependent term vanishes, i.e. we have:
\begin{equation}
    \int_{0}^{2\pi} \sin^{n}\phi d\phi =0,
\end{equation}
Therefore, $G_{zz,n}=0$ for \textit{odd} values of $n$. On the other hand, for \textit{even} values of $n$, we have:
\begin{multline} \label{phi_int_1D}
    \int_{0}^{2\pi} \sin^{n}\phi d\phi = 2B\left(\frac{1}{2},\frac{n}{2}+\frac{1}{2} \right)=\frac{2\Gamma\left(\frac{1}{2}\right)\Gamma\left(\frac{n}{2}+1\right)}{\Gamma\left(\frac{n}{2}+1 \right)}
    \\= \frac{2\sqrt{\pi}}{(\frac{n}{2})!} \left[\frac{n!}{2^{n}(\frac{n}{2})!} \sqrt{\pi} \right]= \frac{\pi}{2^{n-1}} \frac{n!}{\left[(\frac{n}{2})!\right]^{2}},
\end{multline}
with $B$ and $\Gamma$ standing for Beta and Gamma functions respectively \cite{beta_gamma}. Also, when $n$ is even, $n+3$ is \textit{odd}, yielding: 
\begin{multline} \label{theta_int_1D}
    \int_{0}^{\pi} \sin^{n+3}\theta d\theta =B\left(\frac{1}{2},\frac{n}{2}+2\right)=\frac{\Gamma\left(\frac{1}{2}\right)\Gamma\left(\frac{n}{2}+2\right)}{\Gamma\left(\frac{n}{2}+\frac{5}{2} \right)}
    \\=\sqrt{\pi}(\frac{n}{2}+1)! \left[\frac{2^{n+4}(\frac{n}{2}+2)!}{\sqrt{\pi}(n+4)!} \right]
    \\= \frac{\left[(\frac{n}{2})!\right]^{2}}{n!} \left(\frac{n}{2}+1\right) \left[ \frac{2^{n+4}\left(\frac{n}{2}+2\right)\left(\frac{n}{2}+1\right)}{(n+4)(n+3)(n+2)(n+1)}\right]
    \\= \frac{\left[(\frac{n}{2})!\right]^{2}}{n!} \left(\frac{n}{2}+1\right) \left[ \frac{2^{n+2}}{(n+3)(n+1)} \right]
\end{multline}

Therefore, using \eqref{gzz_n}, \eqref{phi_int_1D} and \eqref{theta_int_1D}, we obtain for even values of $n$:
\begin{equation}
    G_{zz,n}=\left(\frac{n}{2}+1\right) W_{n},
\end{equation}
where,
\begin{equation} \label{eq_wn}
    W_{n}=\frac{8\pi}{(n+1)(n+3)}.
\end{equation}
$W_{n}$ is a ``general weighing factor'' for this one-dimensional ID array scenario, which would soon prove to be readily applicable for the more general three dimensional case. The several relevant formulas used to simplify the integrations are collected in the Appendix. 

The rest of the derivations for $G_{pq,n}$ follows a similar route, and consequently will not be elaborately shown here.  
With the help of a symbolic computer package (e.g., the symbolic toolbox of MATLAB or Mathematica), further verification of the detailed  derived expressions for $G_{pq,n}$ were conducted by the authors. The following general observations can be drawn from the results: 
\begin{enumerate}
    \item The following condition holds always true:
    \begin{equation}
    G_{pq,n}=0 \; \text{for odd values of $n$ for all $p,q$.}
    \end{equation}
    This is very significant, since it literally halves the number of integrations to be solved. 
    \item The coefficients for mutually orthogonal ID pairs all vanish, i.e. for all values of $n$:
    \begin{multline}
        G_{xy,n}=G_{yx,n}=G_{yz,n}\\=G_{zy,n}=G_{xz,n}=G_{zx,n}=0.
    \end{multline}
    \item The coefficients for the ID pairs orthogonal to the placement axes (i.e. for $x$-directed or $z$-directed ID pairs) are identical, and can be expressed as:
    \begin{equation}
        G_{xx,n}=G_{zz,n}=\left(\frac{n}{2}+1\right)G_{yy,n}.
    \end{equation}
    where $G_{yy,n}=W_{n}$. 
\end{enumerate}

These analytical results and the various details about the behaviour of various terms indexed by $n$ will be fully exploited in what follows to build efficient and robust cross-correlation computation algorithms for massive MIMO.




\subsection{Analytical CGF Estimation for Two-dimensional Planar Dipole Arrays}
In the last section, we considered that placement of IDs is restricted along $y$-axis, i.e. $x_{d}=z_{d}=0$, which significantly simplified the scenario. Next, let us take the case of a two-dimensional/planar array of IDs placed in the $yz$-plane. Since here $x_{d}=0$, we have $r_{d}=y_{d}\sin\theta\sin\phi + z_{d}\cos\theta$. By deploying the binomial series to expand $r_{d}$ and following some algebraic manipulations, we get:
\begin{multline}
    \left(r_{d}\right)^{n}=\left(y_{d}\sin\theta\sin\phi + z_{d}\cos\theta \right)^{n} \\= \sum_{m=0}^{n} \frac{n!}{m!(n-m)!} y_{d}^{m} z_{d}^{n-m}  \sin^{m}\theta \cos^{n-m}\theta \sin^{m}\phi.
\end{multline}
Therefore, following \eqref{ipq_0}, the expression for $I_{pq,n}$ becomes:
\begin{equation} \label{ipq_yz}
    I_{pq,n}=\frac{1}{n!} \sum_{m=0}^{n} \frac{n!}{m!(n-m)!} V^{n}_{pq,m} y_{d}^{m} z_{d}^{n-m}, 
\end{equation}
where,
\begin{equation} \label{vpq_yz}
    V^{n}_{pq,m} =\int_{0}^{\pi} \int_{0}^{2\pi} f_{pq} \sin^{m}\theta \cos^{n-m}\theta \sin^{m}\phi d\theta d\phi.
\end{equation}
To solve for $V^{n}_{pq,m}$ (which will finally lead to $I_{pq,n}$) we need to carefully choose  $f_{pq}$ expressions from \eqref{fpq} and apply suitable properties of Beta and Gamma functions. We demonstrate the solution for $I_{zz,n}$ here.
\begin{equation}
    V^{n}_{zz,m}=\left[\int_{0}^{\pi} \sin^{m+3}\theta \cos^{n-m}\theta d\theta \right] \left[\int_{0}^{2\pi} \sin^{m}\phi d\phi\right]
\end{equation}
Note that, the integral with $\phi$ vanishes for \textit{odd} values of $m$. Similar to the 1D case, we have for \textit{even} $m$, we have:
\begin{equation} \label{phi_term_2D}
    \int_{0}^{2\pi} \sin^{m}\phi d\phi=2B\left(\frac{1}{2},\frac{m}{2}+\frac{1}{2}\right)=\frac{\pi}{2^{m-1}} \frac{m!}{\left[(\frac{m}{2})!\right]^{2}}.
\end{equation}
Now, we consider the two scenarios of $n$. When $n$ is \textit{odd} with $m$ being \textit{even}, both the quantities $m+3$ and $n-m$ are \textit{odd}. Therefore using the fact that cosine function $\cos\theta$ is odd with respect to $\pi/2$ we have:
\begin{equation}
    \int_{0}^{\pi} \sin^{m+3}\theta \cos^{n-m}\theta d\theta =0.
\end{equation}
Once again, we have $I_{zz,n}=0$ for \textit{odd} values of $n$. On the other hand, when $n$ is \textit{even} with $m$ also being \textit{even}, $m+3$ is \textit{odd} while $n-m$ is \textit{even}. Therefore,  
\begin{multline}  \label{theta_term_2D}
    \int_{0}^{\pi} \sin^{m+3}\theta \cos^{n-m}\theta d\theta = B\left(\frac{n-m}{2}+\frac{1}{2},\frac{m}{2}+2 \right) \\=\frac{\Gamma\left(\frac{n-m}{2}+\frac{1}{2}\right)\Gamma\left(\frac{m}{2}+2 \right)}{\Gamma\left(\frac{n+4}{2}+\frac{1}{2} \right)}
    \\=\left(\frac{m}{2}+1 \right)! \left[\frac{(n-m)!}{2^{n-m}(\frac{n}{2}-\frac{m}{2})!} \right] \left[ \frac{2^{n+4}(\frac{n}{2}+2)!}{(n+4)!} \right].
\end{multline}
Using \eqref{theta_term_2D} and \eqref{phi_term_2D} in \eqref{vpq_yz}, we obtain:
\begin{multline}
    V^{n}_{zz,m}=2^{5}\pi\left[\frac{m!(n-m)!}{n!} \right]\left[\frac{(\frac{n}{2})!}{(\frac{n}{2}-\frac{m}{2})!(\frac{m}{2})!} \right] \\
    \times \left[\frac{\left(\frac{m}{2}+1 \right)\left(\frac{n}{2}+2 \right)\left(\frac{n}{2}+1 \right)}{(n+4)(n+3)(n+2)(n+1)}\right] 
    \\=\frac{8\pi\left(\frac{m}{2}+1 \right)}{(n+3)(n+1)} \left[\frac{m!(n-m)!}{n!} \right] \left[\frac{(\frac{n}{2})!}{(\frac{n}{2}-\frac{m}{2})!(\frac{m}{2})!} \right].  
\end{multline}
When $V^{n}_{zz,m}$ is substituted in \eqref{ipq_yz} and the expression for $W_{n}$ is recognized from \eqref{eq_wn}, the expression for $I_{zz,n}$ becomes:
\begin{equation} \label{izz_n_0}
    I_{zz,n}=\frac{W_{n}}{n!} \sum_{m=0}^{n}  \left[\frac{\left(\frac{m}{2}+1 \right)(\frac{n}{2})!}{(\frac{n}{2}-\frac{m}{2})!(\frac{m}{2})!} \right] y_{d}^{m}z_{d}^{n-m}, 
\end{equation}
At this point, we notice that $I_{zz,n}$ is actually a sum of two series-summations as follows:
\begin{equation}
    \sum_{m=0}^{n} \left[\frac{(\frac{n}{2})!}{(\frac{n}{2}-\frac{m}{2})!(\frac{m}{2})!} \right] y_{d}^{m}z_{d}^{n-m} = (y_{d}^{2}+z_{d}^{2})^{\frac{n}{2}},
\end{equation}
\begin{multline}
    \sum_{m=0}^{n} \left[\frac{(\frac{n}{2})!}{(\frac{n}{2}-\frac{m}{2})!(\frac{m}{2}-1)!} \right] y_{d}^{m}z_{d}^{n-m} 
    \\= y_{d}^{2}\left(\frac{n}{2}\right) \sum_{m=0}^{n} \left[\frac{(\frac{n}{2}-1)!}{(\frac{n}{2}-\frac{m}{2})!(\frac{m}{2}-1)!} \right] y_{d}^{m-2}z_{d}^{n-m}
    \\=\frac{n}{2}y_{d}^{2} (y_{d}^{2}+z_{d}^{2})^{\frac{n}{2}-1}.
\end{multline}
After substituting these series summation values in \eqref{izz_n_0}, the final expression for $I_{zz,n}$ reduces to:
\begin{equation}
    I_{zz,n} = \frac{W_{n}}{n!} \left(y_{d}^{2}+z_{d}^{2} \right)^{\left(\frac{n}{2}-1\right)}\left[\left(\frac{n}{2}+1\right)y_{d}^{2}+z_{d}^{2} \right].
\end{equation}
In a similar fashion, the expressions for other $I_{pq,n}$ for \textit{even} values of $n$ can be derived as follows:
\begin{equation}
    I_{xx,n} = \frac{W_{n}}{n!} \left( \frac{n}{2}+1 \right)\left(y_{d}^{2}+z_{d}^{2} \right)^{\left(n-1\right)},
\end{equation}
\begin{equation}
    I_{yy,n} = \frac{W_{n}}{n!} \left(y_{d}^{2}+z_{d}^{2} \right)^{\left(\frac{n}{2}-1\right)}\left[y_{d}^{2}+\left(\frac{n}{2}+1\right)z_{d}^{2} \right],
\end{equation}
\begin{equation}
    I_{yz,n} = I_{zy,n} = -\frac{W_{n}}{n!} \left( \frac{n}{2} \right) \left(y_{d}z_{d}\right) \left(y_{d}^{2}+z_{d}^{2} \right)^{\left(\frac{n}{2}-1\right)} 
\end{equation}
\begin{equation} \label{iorth_yz}
I_{xy,n} = I_{yx,n} = I_{xz,n} = I_{zx,n} = 0. 
\end{equation}
Note that, the condition $I_{pq,n}=0$ for \textit{odd} values of $n$ holds true. Furthermore, \eqref{iorth_yz} suggests that the IDs oriented orthogonal to the plane of arrangement (i.e. $x$-directed IDs) do not have any correlation with the IDs oriented along the plane of arrangement (i.e. $y$-directed or $z$-directed IDs).

\subsection{Analytical CGF Estimation for Three-Dimensional Dipole Arrays: Generalized Case}
Finally we consider the most general case of three-dimensional arrays with no restrictions imposed on the dipole locations, i.e. in general, $x_{d} \neq y_{d} \neq z_{d} \neq 0$. Here, it turns out we have to deal with a trinomial expansion or successive binomial expansions of $(r_{d})^{n}$ where 
\begin{equation}
    r_{d}=x_{d}\sin\theta\cos\phi+y_{d}\sin\theta\sin\phi + z_{d}\cos\theta. 
\end{equation}
Therefore, the analytical integrations needed to evaluate $I_{pq,n}$ (see \eqref{ipq_0}) become slightly more complicated. Performing integration both by-hand using the properties of Beta and Gamma functions as before, we determine the following general formula for $I_{pq,n}$ for \textit{even} values of $n$:   
\begin{equation} \label{ixx_n}
    I_{xx,n} = \frac{W_{n}}{n!} \left[x_{d}^{2}+\left( \frac{n}{2}+1 \right)y_{d}^{2}+\left( \frac{n}{2}+1 \right)z_{d}^{2} \right] d^{n-2},
\end{equation}
\begin{equation} \label{iyy_n}
    I_{yy,n} = \frac{W_{n}}{n!} \left[\left( \frac{n}{2}+1 \right)x_{d}^{2}+y_{d}^{2}+\left( \frac{n}{2}+1 \right)z_{d}^{2} \right] d^{n-2},
\end{equation}
\begin{equation} \label{izz_n}
    I_{zz,n} = \frac{W_{n}}{n!} \left[\left( \frac{n}{2}+1 \right)x_{d}^{2}+\left( \frac{n}{2}+1 \right)y_{d}^{2}+z_{d}^{2} \right] d^{n-2},
\end{equation}
\begin{equation} \label{ixy_n}
    I_{xy,n}=I_{yx,n}=-\frac{W_{n}}{n!} \left[ \frac{nx_{d}y_{d}}{2} \right] d^{n-2},
\end{equation}
\begin{equation} \label{iyz_n}
    I_{yz,n}=I_{zy,n}=-\frac{W_{n}}{n!} \left[ \frac{ny_{d}z_{d}}{2} \right] d^{n-2},  
\end{equation}
\begin{equation} \label{izx_n}
    I_{zx,n}=I_{xz,n}=-\frac{W_{n}}{n!} \left[ \frac{nx_{d}z_{d}}{2} \right] d^{n-2},  
\end{equation}
where,
\begin{equation}  \label{d_xdydzd}
    d=\sqrt{x_{d}^{2}+y_{d}^{2}+z_{d}^{2}}
\end{equation}
However the results are further validated by use of the symbolic toolbox in MATLAB (see appendix). It is observed that for \textit{odd} values of $n$, $I_{pq,n}=0$. Also note that, the expressions for one-dimensional and two-dimensional ID arrays can be easily computed back from \eqref{ixx_n}-\eqref{izx_n}, substituting $x_{d},y_{d}$ and $z_{d}$ accordingly. \textit{Consequently, the results of this subsection \eqref{ixx_n}-\eqref{izz_n} are the most general} but we opted for presenting the one- and two-dimensional cases for convenience since the mathematical treatment is considerably more complex in three dimensional arrays while lower-dimensional MIMO systems tend to be more commonly encountered in practice. 

Now, similar to our approach for the linear array (or 1D case), it is crucial to predict the maximum number of terms $N$ needed to truncate the series in \eqref{cpq_approx}. With that objective in mind, we start by carefully examining $I_{xx},n$, where we note that for $n>0$:
\begin{multline}
    I_{xx,n}=\frac{W_{n}}{n!} \left[x_{d}^{2}+\left( \frac{n}{2}+1 \right)y_{d}^{2}+\left( \frac{n}{2}+1 \right)z_{d}^{2} \right] d^{n-2} \\<\frac{W_{n}}{n!}\left( \frac{n}{2}+1 \right)\left[x_{d}^{2}+y_{d}^{2}+z_{d}^{2} \right]d^{n-2}.
\end{multline}
Using $d$ from \eqref{d_xdydzd}, following the same procedure for all all $p=x,y,z$, and applying the Stirling's approximation \eqref{stirling}, the upper-bound of $I_{pp,n}$ can be expressed as:
\begin{equation}
    I_{pp,n}|_{\text{U.B.}}= W_{n} \left( \frac{n}{2}+1 \right) \frac{d^{n}}{n!} = W_{n} \left( \frac{n+2}{2\sqrt{2\pi n}} \right)  \left(\frac{ep\lambda}{n}\right)^{n},
\end{equation}
where $d=p\lambda$. Therefore, when this $I_{pp,n}$ is used in \eqref{cpq_approx} containing the $(jk)^{n}$ term where $k=2\pi/\lambda$, we would obtain a term $(2\pi ep /n)^{n}$, very much similar to \eqref{cpq_approx_1D}. Therefore, it can be deduced that the maximum $N$-value for the three-dimensional array situation also follows the same guideline given by \eqref{ruleofthumb}.

\section{Conclusion}
The present paper analytically estimates the frequency-domain CGF tensor by employing an accurate series expansion, followed by detailed integration of functions involving trigonemetric expressions of elevation ($\theta$) and azimuth ($\phi$) angles. Expressions for coefficients $I_{pq,n}$ are formulated for the general three-dimensional case (see \eqref{ixx_n}-\eqref{izz_n}), which enables very efficient CGF-tensor calculation by simply using the knowledge of inter-element spacing ($x_{d}$, $y_{d}$ and $z_{d}$) between the ID-pairs, along with their relative polarizations. We systematically demonstrate the effect of number of terms $N$ needed to truncate the series used for approximating the CGFs (see \eqref{cpq_approx}), and proposed an effective equation \eqref{ruleofthumb} for estimating $N$ from spacing between the IDs. 


\section*{Appendix}
While deriving the $I_{pq,n}$ expressions for the one/two/three dimensional cases, we encountered several definite integrations (over the full $\theta$-$\phi$ space), involving arbitrary powers of trigonometric functions $\sin\theta$, $\cos\theta$, $\sin\phi$ and $\cos\phi$. To obtained closed form solutions of these integrations, we utilized the well-known Beta functions $B(m,n)$ having the mathematical form \cite{beta_gamma}
\begin{equation}
    B(m,n)=\int_{0}^{1} z^{m-1} (1-z)^{n-1} dz. 
\end{equation}
The connection between the Beta functions and the integrals involving trigonometric functions is established using
\begin{equation}
    B(m+1,n+1)=2\int_{0}^{\frac{\pi}{2}} \cos^{2m+1}\theta \sin^{2n+1}\theta d\theta. 
\end{equation}
Now, the analytical evaluation of Beta functions is performed by invoking the Gamma functions via the relation \cite{beta_gamma}
\begin{equation}
    B(m,n)=\frac{\Gamma(m)\Gamma(n)}{\Gamma(m+n)},
\end{equation}
where,
\begin{equation}
    \Gamma\left(n+1\right)=n! \;\; \text{and} \;\; \Gamma\left(n+\frac{1}{2}\right) = \frac{(2n)!}{2^{2n} n!} \sqrt{\pi}.
\end{equation}

Next, we provide solutions for the \textit{four} general classes of definite integrals involving $\sin\theta$ and $\cos\theta$, that are encountered during the derivations.
\vspace{5pt}

\textit{Class-I: Power of $sin$ odd, Power of $cos$ even:}
\begin{equation}
    \int_{0}^{\pi} \sin^{2m+1}\theta \cos^{2n}\theta d\theta = B\left(n+\frac{1}{2},m+1\right).
\end{equation}
\begin{equation}
\int_{0}^{2\pi} \sin^{2m+1}\theta \cos^{2n}\theta d\theta = 0. 
\end{equation}

\vspace{5pt}

\textit{Class-II: Power of $sin$ odd, Power of $cos$ odd:}
\begin{equation}
    \int_{0}^{\pi} \sin^{2m+1}\theta \cos^{2n+1}\theta d\theta = \int_{0}^{2\pi} \sin^{2m+1}\theta \cos^{2n+1}\theta d\theta =  0.
\end{equation}

\vspace{5pt}

\textit{Class-III: Power of $sin$ even, Power of $cos$ odd:}
\begin{equation}
    \int_{0}^{\pi} \sin^{2m}\theta \cos^{2n+1}\theta d\theta = \int_{0}^{2\pi} \sin^{2m}\theta \cos^{2n+1}\theta d\theta = 0.
\end{equation}

\vspace{5pt}

\textit{Class-IV: Power of $sin$ even, Power of $cos$ even:}
\begin{multline}
    \int_{0}^{\pi} \sin^{2m}\theta \cos^{2n}\theta d\theta=\frac{1}{2}\int_{0}^{2\pi} \sin^{2m}\theta \cos^{2n}\theta d\theta\\=B\left(n+\frac{1}{2},m+\frac{1}{2}\right). 
\end{multline}

\end{document}